\documentclass[manuscript]{aastex}
\usepackage{natbib}
\usepackage{amsfonts}
\usepackage{graphicx}

\slugcomment{}

\shorttitle{Polarimetry of DG Tau at 350\,$\mu$m}
\shortauthors{Krejny et al.}

\begin{document}


\title{Polarimetry of DG Tau at 350\,$\mu$m}


\author{M. Krejny\altaffilmark{*}, T. Matthews, and G. Novak}
\affil{Northwestern University, Department of Physics and Astronomy, 2131 Tech Dr., Evanston, IL 60628, USA}

\author{J. Cho}
\affil{Chungnam National University, Dept. of Astronomy and Space Science, Daejeon, South Korea}

\author{H. Li}
\affil{Harvard Center for Astrophysics, 60 Garden Street, Cambridge, MA 02138, USA}

\author{H. Shinnaga}
\affil{Caltech Submillimeter Observatory, 111 Nowelo St., Hilo, HI 96720, USA}

\and

\author{J.~E. Vaillancourt}
\affil{California Institute of Technology, Division of Physics, Mathematics, and Astronomy, 1200 E. California Blvd., Pasadena, CA 91125, USA}


\altaffiltext{*}{Current address:  Department of Astronomy, University of Minnesota, 116 Church St. SE, Minneapolis, MN 55455, USA}


\begin{abstract}
We present the first 350\,$\mu$m polarization measurement for the disk of
the T Tauri star (TTS) DG Tau.  The data were obtained using the SHARP
polarimeter at the Caltech Submillimeter Observatory.  We did not detect
polarization but we provide constraints on the normalized Stokes
parameters $q$ and $u$.  We derive information about the polarization 
spectrum by comparing our 350\,$\mu$m result with an 850\,$\mu$m
polarization detection previously published for this source.
This empirical information concerning the polarization spectrum 
disagrees with the predictions of a recent model for TTS disk
polarization.  We conclude, however, that adding more mass to the model
disk would probably result in model polarization spectra that agree much better 
with the 350 and 850\,$\mu$m polarimetric observations.  We suggest that multiwavelength
polarimetry of TTS disk emission may provide a promising method for
probing the opacity of TTS disks.

\end{abstract}



\section{INTRODUCTION}

T Tauri Stars (TTSs) are low-mass Young Stellar Objects, i.e., objects that have masses similar to that of the Sun and have not yet reached the main sequence.  They are characterized primarily by their large infrared excesses that are caused by the dust in their circumstellar disks.  TTSs are being studied to learn more about possible first steps in planet formation, e.g., dust coagulation from interstellar medium (ISM) sizes ($< 1\,\mu$m) up to 1\,mm.  Evidence for grain size growth is found in numerous studies of the millimeter and submillimeter spectral energy distributions (SEDs) of TTSs \citep[e.g.,][]{beckwith91, rodmann06, draine06}.  In these studies,  the mass opacity index $\beta$ for TTS disks is found to be generally less than 2, indicative of grain sizes larger than those in the ISM.  However, the methods used to determine $\beta$ require assumptions about the optical depths of the disks.  Due to model degeneracies, the results obtained for the grain size, though compelling, are not fully conclusive.

Another property of TTS disk emission that can be observed, besides its SED, is its state of linear polarization.   In general, dust grains can produce polarization by any of three mechanisms:  thermal emission, selective extinction, and/or scattering.  \citet{hildebrand88} reviews observations of polarized thermal emission from magnetically aligned interstellar dust grains at far-infrared and submillimeter wavelengths, and shows that one can generally assume that the clouds are optically thin and that selective extinction and scattering are thus negligible \citep[see also][]{novak89,lazarian07}.  The principal reason for this is that the grain size $a$ is much smaller than the wavelength $\lambda$ for these observations.  At shorter wavelengths, magnetically aligned dust grains can  produce polarization by selective extinction of starlight, as first seen by \citet{hiltner49} and \citet{hall49} in the optical, and later observed by others in the near infrared \citep[e.g.,][]{jones92}.  Selective extinction  (specifically, selective absorption) has also been seen at far-infrared wavelengths in the unusually dense core of Sagittarius B2 \citep{dowell97}.  For polarization by scattering, aligned grains are not needed.  This mechanism can sometimes produce large polarizations, and has been observed in the optical and NIR \citep[e.g.,][]{werner83}.  

If the dust grains in a TTS disk are aligned, then one should observe polarized millimeter and submillimeter thermal emission.  Also, if the grains are growing up to millimeter sizes, then polarization by selective extinction and/or scattering may become non-negligible at millimeter and submillimeter wavelengths.  To see why, note that the submillimeter opacity can exceed unity for some large-grain models created for protostellar disks \citep{dalessio01}.  



\citet{tamura99} were the first to detect polarization in TTS disks in the submillimeter; they detected 850\,$\mu$m polarization in two approximately edge-on TTS disks.  In both cases, the measured polarization magnitude was found to be about 3\% and the measured polarization angle was orthogonal to the plane of the disk.  They assumed that the polarization mechanism was thermal emission by magnetically aligned dust grains, and on this basis they inferred that the magnetic field orientation for each disk was parallel to the plane of the disk, implying a toroidal magnetic field structure.  However, \citet{tamura99} did not consider alternative polarization 
mechanisms, so their conclusions about disk magnetic fields must be considered preliminary.  Because the cross sections for scattering and absorption can depend strongly on $(a/\lambda)$, measurements of the polarization spectrum of the disk emission may constrain the polarization mechanism.  This, in turn, may lead to definitive information about the magnetic field geometry and/or new constraints on the grain size distribution.


In this paper, we present the first 350\,$\mu$m polarimetric observations of DG Tau, one of the two TTSs studied by \citet{tamura99} at 850\,$\mu$m.  The observations were made at the Caltech Submillimeter Observatory (CSO) using SHARP, the SHARC-II Polarimeter \citep{li08}.  In \S\,2, we describe the observations and the analysis, and we present our results.  In \S\,3 we discuss a model for TTS disk polarization that has recently been published by \citet{cho07}, and we compare the predictions of this model with initial empirical information on the polarization spectrum of DG Tau derived by comparing our 350\,$\mu$m measurement with the 850\,$\mu$m measurement of \citet{tamura99}.  

\section{Observations, Analysis and Results}

DG Tau is located in the Taurus-Auriga star forming cloud, at a distance of 140\,pc \citep{beckwith90}.  It has a mass of 0.6\,M$_{\odot}$ and luminosity of 1.7\,L$_{\odot}$ \citep{cohen79, beckwith90}.  It is a younger TTS, with an age of approximately 3 $\times$ 10$^{5}$\,yr \citep{beckwith90}.  DG Tau has a prominent optical jet at a position angle (measured from north to east) of 226$^{\circ}$  \citep{mundt83}, and shows forbidden-line emission, making it a classical TTS.  

Disk models created to reproduce the SED suggest a disk mass of 0.03\,M$_{\odot}$ \citep{adams90}.  Based on 2\,mm observations, the dust disk is estimated to have a radius of 109 $\pm$ 22\,AU and an inclination angle of 70$^{\circ}$, with 90$^{\circ}$ signifying an edge-on disk \citep{kitamura96}.  The dust disk is \textit{not} orthogonal to the outflow axis; it is misaligned by almost 30$^{\circ}$.  There is also a large $^{13}$CO gas disk with a major axis of 2800\,AU, perpendicular to the optical jet \citep{sargent89, kitamura96}.

SHARP is a dual-polarization fore-optics module that adds polarimetric capability to the CSO's SHARC-II camera \citep{li08}.  The angular resolution of SHARP at 350\,$\mu$m is $\sim 9^{\prime\prime}$, so the disk of DG Tau is not resolved.  Observations were made during 2007 February 12-14, August 9, and August 13.  Table 1 lists observing information.  We used chop-nod mode for these observations \citep{li08, hildebrand00}.  Data were collected in groups of dithers, with one dither consisting of four half-wave plate (HWP) cycles \citep{hildebrand00} and each cycle having a slightly different pointing offset.  Each cycle had a duration of eight minutes, with data collected at four HWP angles:  0$^{\circ}$, 22.5$^{\circ}$, 45$^{\circ}$, and 67.5$^{\circ}$.  

We used the (in-house designed) software programs \texttt{SHARPINTEG} and \texttt{SHARPCOMBINE} to analyze the data.  \texttt{SHARPINTEG} processes the individual cycles to yield maps of the Stokes Parameters $I$, $Q$, and $U$.  \texttt{SHARPCOMBINE} accepts these maps as inputs and interpolates them onto its own regular equatorial-coordinate grid, with pixel sizes that are half the size of the SHARC-II detector pixels (one ``plot pixel'' = $\sim 2.3^{\prime\prime} \times 2.3^{\prime\prime}$). The principles used in this interpolation are decribed by \citet{houde07}.  Interpolation is necessary because of the dithering and sky rotation \citep{li08}, and entails a modest loss of angular resolution.  For this process, we used a smoothing kernel having a Gaussian shape of 9$^{\prime\prime}$ FWHM.  The kernel's cut-off diameter was also set to 9$^{\prime\prime}$.  


The above procedures have proven reliable in past work, but residual ``sky noise'' (Hildebrand et al.\ 2000) or other variations in the radiative load can in principle cause spurious additive signals in both total and polarized flux.  Because our target is unresolved by SHARP, our field of view 
consists mostly of blank sky for which $I$, $Q$, and $U$ should be zero.  We have taken advantage of this to remove the kinds of spurious 
signals described above, by performing the equivalent of synthetic aperture photometry \citep{howell00} on the $I$, $Q$, and $U$ maps that 
are produced by \texttt{SHARPCOMBINE}.  The effect of this process is to remove any ``DC offsets'' in flux and/or polarized flux that are uniform 
across the array. For each of the three maps ($I$, $Q$, and $U$), we defined an annulus centered on the central pixel of the 
map (the location of DG Tau), calculated a (straight) average over the annulus pixels, and then subtracted this average value (the DC offset) from the value determined at the center pixel.  The inner radius of the annulus was chosen to be twice the kernel cut-off radius; this ensured that \texttt{SHARPINTEG} output values used to compute the central \texttt{SHARPCOMBINE} output pixel would not also be used to compute the DC offset, and visa versa.  The outer radius was chosen to give a total number of annulus pixels that was 10 times that of the kernel; larger choices resulted in inclusion of parts of the \texttt{SHARPCOMBINE} maps having lower sampling, resulting in larger noise levels.  The annulus had a total sky coverage corresponding approximately to that of one instantaneous pointing of the 12$\times$12 pixel ($55^{\prime\prime} \times 55^{\prime\prime}$) polarimetry array of SHARP.  Our use of synthetic aperture photometry is illustrated in Figure 1.

From the final values of $I$, $Q$, and $U$ for DG Tau, and the associated
statistical errors, normalized Stokes parameters ($q = Q/I$ and $u = U/I$) were then calculated to obtain our final result.  Note that we also apply the usual corrections for instrumental polarization and polarimetric efficiency, as detailed in \citet{li08}.  Note that in \texttt{SHARPINTEG} and \texttt{SHARPCOMBINE}, errors are propagated from theshort-time-scale errors determined from each nod, as described by \citet{hildebrand00}, \citet{kirby05}, and references therein.  In order to assess the effects of systematic error on our results, we processed the data a second time, but with the data grouped into seven temporal bins.  \texttt{SHARPCOMBINE} $I$, $Q$, and $U$ maps were created separately for each bin, and DC values derived from annuli were subtracted from the $I$, $Q$, $U$ values measured at the centers of these maps, for each of the bins.   The purpose of the binning was to determine the reduced chi squared ($\chi^{2}_{red}$) of the $q$ and $u$ data, which may be compared to that obtained without the DC offset subtraction.  We found that the offset subtraction lowered $\chi^{2}_{red}$ from 1.5 and 1.6  to 0.9 and 0.8 (for $q$ and $u$, respectively).  Thus, after the DC offset subtraction, the data appear to be free from systematic error.  We found that the $q$ and $u$ values obtained via the two methods (averaging of all files in one bin vs.\ using seven bins and taking a weighted average over results for individual bins) are very consistent.  We adopt the values from the first method as our final result, which are $q = -0.0086 \pm 0.0060 $ and $u = -0.0012 \pm 0.0061$.   

Finally, we note that in examining the $I$ maps we found a small amount of flux from DG Tau, $\sim$ 2\%, contaminating the annulus (see upper right panel in Figure 1).  However, this flux should be consistent across all Stokes parameters.  If 0.98 of the flux for each Stokes parameter remains after the DC offset subtraction, then this factor should cancel out upon determination of the normalized Stokes parameters $q$ and $u$.

\section{DISCUSSION}

\subsection{The Model of Cho and Lazarian}\label{sec:cl}

\citet{cho07} present a model for polarized thermal emission from magnetically aligned grains in a TTS disk.  Besides treating polarized emission, their model naturally incorporates the effects of large absorption optical depths.  The model does not include the effects of polarization by 
scattering.  Instead, they provide a simple set of approximate calculations which indicate that scattering should not be dominant in the submillimeter. 


\cite{cho07} use a flared, two-layered (surface and interior) disk model, incorporating a distribution of grain sizes extending up to 1\,mm.  The disk is permeated by a regular, toroidal magnetic field, and the model predicts that the submillimeter/millimeter emission should be polarized in a direction orthogonal to this field, giving polarization perpendicular to the disk plane for a TTS disk viewed approximately edge-on.

\citet{cho07} determine the degree of grain alignment in 
the disk, under the assumption that grains are brought into alignment 
with the magnetic field via the radiative torque mechanism.  The resulting degree of alignment depends on the grain size, the local 
radiation environment, and the local gas density.  Grains of size $a$ are 
most easily aligned by radiation having wavelength comparable to $a$.  One 
result of the \citet{cho07} model is that grain alignment is generally 
better at larger distances from the star.  Also, large grains are 
generally better aligned than small grains.

In the \citet{cho07} model, many factors affect the degree of 
polarization of the grains' emission, including of course the degree of 
grain alignment and the axis ratio of the grains (assumed to be oblate 
spheroids).  One additional factor that will turn out to be important for our 
comparisons with observations in \S\,\ref{sec:model} is the absorption optical depth, which for some wavelengths 
and sight-lines can approach or exceed unity.  In these cases, the calculated polarization 
will tend to be suppressed.  This is because the polarization by absorption that the grains near the front of the disk impress upon the 
radiation from the grains near the back of the disk is orthogonal to the intrinsic polarization of that radiation.  This effect is also discussed 
by \citet{hildebrand00}, and we shall refer to it as \textit{polarization self suppression}, or PSS.  



In the \citet{cho07} model, the disk interior becomes optically thick for $\lambda < $ 100\,$\mu$m, while the surface layer remains optically thin.  The hotter surface grains dominate the mid-IR polarization, while the cooler interior dust grains dominate polarization in the far-infrared/submillimeter.  Also, grains located further from the star tend to dominate the long wavelength polarization.

In their Figure 12, \citet{cho07} show predicted polarization spectra for unresolved disks at various inclination angles $i$.  Note that as $i$ approaches zero (face-on view), the polarization also approaches zero, for all wavelengths.  This is due to polarization cancellation, as the disk's polarization pattern becomes radial for $i = 0^{\circ}$





 Finally, we review the discussion of polarization by scattering in \citet{cho07}.  For ISM sized grains, far-infrared/subillimeter scattering cross sections are negligible, but this is not necessarily true for large, 1000\,$\mu$m grains.  \citet{cho07} carried out a simple calculation to estimate the relative importance of polarization by scattering vs. polarization by thermal emission.  They chose a point in the midplane of the disk, and calculated both the scattered flux $F_{sca}$ and the flux from the thermally emitting dust grains, $F_{em}$.  The polarized flux due to scattering is proportional to the first of these two quantities while that due to thermal emission is proportional to the second.  Figure 16 of \citet{cho07} plots the ratio $F_{sca}/F_{em}$ as a function of radial distance from the central star.  For submillimeter wavelengths, the ratio is less than unity, except at small radial distances.  Since the submillimeter polarization is dominated by the outer part of the disk, the authors  conclude that polarization by scattering is not dominant at these long wavelengths.

\subsection{Comparing the Observations to Model Predictions}\label{sec:model}

The $q$ and $u$ values for the measured 350\,$\mu$m polarization (see \S\,2) give a percent polarization of $P_{350\,\mu m} = (0.9 \pm 0.6)$\%. The degree of polarization reported by \cite{tamura99} at their longer wavelength is $P_{850\,\mu m} = (2.95 \pm 0.89)$\%.  These two values of $P$ agree within 2$\sigma$.  However, Figure \ref{fig:dgtaupol}  plots the 850\,$\mu$m and 350\,$\mu$m measurements in Stokes space, with circles that denote 1 and 2\,$\sigma$ error bars, and it can be seen that the Stokes parameters corresponding to the measurements at the two wavelengths do not agree within 2\,$\sigma$;  the 2\,$\sigma$ error-circles do not overlap.  Thus, the two measurements taken together 
imply significant structure in the polarization spectrum.

The vertically hatched portion of the Stokes plot in Figure\,2 represents the locus of points that are consistent with polarization oriented orthogonal to the plane of the disk within the errors given by \citet{kitamura96}.  For this purpose, the position angle of the disk axis was taken to be $99^{\circ} \pm 10^{\circ}$ \citep{kitamura96}.  As we noted in \S\,1, the 850\,$\mu$m point is consistent with polarization orthogonal to the disk.  Our 350\,$\mu$m measurement is barely consistent with this polarization orientation, and is consistent within 2$\sigma$ with zero.  If we assert, following \citet{tamura99} and \citet{cho07}  that the submillimeter polarization is oriented perpendicular to the plane of the disk, then the measurements shown in Figure\,2 indicate that the polarization must drop by at least a factor of two or three as one moves from 850\,$\mu$m to 350\,$\mu$m.


We compare this result to the submillimeter polarization spectrum predicted by \citet{cho07} for an unresolved disk viewed at a 60$^{\circ}$ inclination angle, which is plotted in the upper right panel of their Figure 12.  (This inclination angle is the plotted angle closest to that of DG Tau; see \S\,2.)  As can be seen in this polarization spectrum, the degree of polarization is reasonably flat across the submillimeter bands; from the original numerical data used by \citet{cho07} to make this plot we find that the ratio of 850 to 350\,$\mu$m polarization $(P_{850}/P_{350})$ is 1.28.  This disagrees with the observations.



One possible difference between the model disk and the actual DG Tau disk is the mass; the model disk's mass is $0.014\,M_{\odot}$, while the mass of the DG Tau disk has been estimated to lie in the range 0.02--0.06\,$M_{\odot}$ \citep{kitamura96, beckwith91, mannings94}.  Increasing the mass of a disk increases the optical depth along any line of sight through the disk.  Recall that large optical depth can reduce polarization via the PSS effect (\S\,\ref{sec:cl}).   Is it possible that the difference between the model disk's polarization spectrum and that of the real DG Tau disk is due to 
suppression of the 350\,$\mu$m polarization in the real disk via a PSS effect induced by the extra optical depth?   To answer this question conclusively, we would have to redo the theoretical work, exploring larger mass values.  However, a preliminary answer can be obtained via extrapolations based on information extracted from \citet{cho07}.  Specifically, we will explore the role of optical depth and PSS in their 0.014 $M_{\odot}$ disk.



First note, however, that the PSS effect must be more severe at shorter wavelengths if it is to cause the observed drop in polarization 
moving from 850\,$\mu$m to 350\,$\mu$m. The frequency dependence of the mass opacity $\kappa_{\nu}$, which is defined as the optical depth $\tau$ divided by the mass column density, is generally taken to be $\kappa_{\nu} \propto \nu^{\beta}$ \citep[e.g., see][]{stahler04}.  Since the mass column density has no wavelength dependence, we must have $\tau \propto \nu^{\beta}$ as well.  For TTS disks, $\beta$ is usually less than 2 in the submillimeter but is rarely negative (see references given in \S\,1).  This implies that, for a given line of sight and a given dust distribution, $\tau$ decreases with increasing wavelength. The same is true for the model 
disk:  The $\kappa_{\nu}$ values employed by \citet{cho07} fall as the wavelength increases.  Thus, the PSS effect should indeed be more significant for our 350\,$\mu$m measurement than for 850\,$\mu$m. 


Turning to the question of the role of optical depth and PSS in the 0.014\,M$_{\odot}$ disk of \citet{cho07}, we next consider the right-hand panel of their Figure 13, which plots the degree of polarization versus inclination angle for a selection of far-infrared/submillimeter wavelengths, again for the case of spatially unresolved measurements.  These plots only consider polarization for the disk interior; for now we shall neglect the surface layers.  From the original numerical data used by \citet{cho07}, we find that the ratio of 100 and 850\,$\mu$m polarization at an inclination $i = 70^{\circ}$ is  $(P_{100}/P_{850}) = 0.15$.   It is reasonable to ask whether this large drop in polarization from 850\,$\mu$m to 100\,$\mu$m is caused by PSS, or is an intrinsic feature of the disk which could result from the fact that the grains that produce polarization at the shorter wavelength reside in warmer regions closer to the star, and are thus less aligned (see \S\,3.1).

This question is resolved when we compare the above polarization ratio for $i = 70^{\circ}$ to values obtained for lower inclination angles.  For $i = 30^{\circ}$, the data 
used to make Figure 13 of \citet{cho07} give $(P_{100}/P_{850}) = 0.39$.  Since the polarization pattern observed becomes centrosymmetric for small inclination angles, the polarization tends to cancel as $i$ approaches zero.  This is a purely geometric effect, unrelated to optical depth, so this polarization cancellation should affect the polarization observed at all wavelengths equally, and the polarization ratio should therefore not change with inclination angle.  However, we have seen that $(P_{100\,\mu m}/P_{850\,\mu m})$ changes very significantly with inclination.  This is in accord with the expected behavior of PSS, because (a) as one increases the optical depth along the line of sight by increasing $i$ the PSS should get stronger, and (b) the PSS should have a more significant effect at 100\,$\mu$m than at 850\,$\mu$m.  We conclude that the dramatic change in polarization between 100\,$\mu$m and 850\,$\mu$m for $i = 70^{\circ}$ (for the disk interior) is due in large part to PSS.  The effect of PSS should not be as dramatic at longer wavelengths, which is confirmed by studying the (350\,$\mu$m/850\,$\mu$m) polarization ratio; at $i = 70^{\circ}$ the ratio is 0.72, and it only increases to 0.83 at $i = 30^{\circ}$.   

As discussed above, the disk mass used by Cho and Lazarian differs from the presumed DG Tau value by a factor of 1.4--4.0.  If one were to repeat the work of \citet{cho07} using a disk mass that is higher than the one they used by a factor in this range then the effects of PSS would certainly 
get worse, and the dramatic PSS effects that we see in the 
$0.014\,M_{\odot}$ disk at 100\,$\mu$m might begin to move into the 
submillimeter.  It seems plausible that the value of the ratio 
$(P_{350}/P_{850})$ for the $i = 70^{\circ}$ case (this value of $i$ 
matches the actual DG Tau disk; see \S~2) could then decrease from 0.72, 
which was found for the $0.014\,M_{\odot}$ disk, to a value below 0.5 
which would match the observations.  

Note that in the above arguments, we have neglected the surface layers, which do emit polarized light.  However, for submillimeter wavelengths ($\lambda > 350$\,$\mu$m) the surface layers do not contribute significantly to the total polarization.  Thus, to understand the 
submillimeter data discussed here we have only considered the polarization properties of the disk interior, as well as how these might be altered by the more severe PSS effect that would accompany a higher mass.

Finally, we consider an alternative explanation for the discrepancy between the predicted and observed polarization spectra, namely, polarization by scattering.  As we discussed in \S\,3.1, \citet{cho07} argue that scattering is less important than polarized 
emission in the submillimeter.  However, their argument is based on a rough 
estimate.  Furthermore, while they do find that the ratio $(F_{sca}/F_{em})$ is less than unity for radii 
corresponding to the outer disk, where the submillimeter radiation originates, 
this ratio nonetheless does hover near 0.5 over much of the relevant 
range.  A disk having a larger mass will have scattering optical depths 
exceeding unity out to relatively larger disk radii, in comparison with 
a lower-mass disk.  Thus, it is likely that if one were to repeat the 
analysis given in \citet{cho07}, but using a larger-mass disk, 
the importance of scattering would be found to be greater.  Would the 
scattering mechanism give a polarization that falls sharply as one moves 
shortward in wavelength from 850 to 350\,$\mu$m?  \citet{cho07} do not evaluate the spectrum for this polarization 
mechanism, but crude toy models by \citet{krejny08} suggest that strong 
features in the polarization spectrum can arise from the wavelength 
dependence of the scattering optical depth.

In summary, when we compare our 350\,$\mu$m polarization measurement to 
the 850\,$\mu$m measurement of \citet{tamura99}, we see evidence for 
significant structure in the polarization spectrum.  This is 
inconsistent with the model of \citet{cho07} which gives 
polarization perpendicular to the disk, with similar magnitudes at the 
two wavelengths.  We have proposed two possible explanations for the 
discrepancy, both of which rely on the fact that the real DG Tau disk is 
likely to have more mass than the model disk.  The first explanation is PSS at 350\,$\mu$m due to the larger absorption optical depth of a more massive 
disk, and the second is the onset of polarization by scattering, 
considered negligible by \citet{cho07} for their lower-mass disk.

If either of our proposed explanations is correct, then the submillimeter 
polarization spectrum of DG Tau is determined by optical depth effects; 
either absorption optical depth or scattering optical depth.  In this 
case, future multi-wavelength far-IR/submm/mm polarimetry of TTS disk 
emission, e.g. with SOFIA, ALMA, and EVLA, may constrain the optical 
depth which is an important unknown in TTS research.

\section*{ACKNOWLEDGEMENTS}

We would like to thank A. Lazarian and B. Whitney for 
their comments and suggestions.  For help with the development and 
commissioning of SHARP, we are grateful to M. Attard, C. D. Dowell, R. 
Hildebrand, M. Houde, L. Kirby, and L. Leeuw.  This work was 
supported by the NSF via grants AST 02-41356 and AST 05-05230 to 
Northwestern University.  Additional support came in the form of a NASA 
GSRP Award to MK.  The CSO is supported by NSF grant AST 05-40882.  

{\it Facilities:} \facility{CSO (SHARP)}

\begin{table}
\center{Table 1:  SHARP 350\,$\mu$m Observations.\label{tab:dgtobs}}
\begin{center}
\begin{tabular}{ccc}
\hline
\textbf{Date} & \textbf{Number of Cycles} & \textbf{$\tau_{225\,GHz}^{*}$} \\
\hline
2007 Feb 12& 11 & 0.033-0.035 \\
2007 Feb 13& 26 & 0.056-0.066 \\
2007 Feb 14& 14 & 0.062-0.074 \\
2007 Aug 9& 4 & 0.065-0.066 \\
2007 Aug 13& 16 & 0.047-0.049\\
\hline
\end{tabular}
\end{center}
\center{\footnotesize{* Zenith atmospheric optical depth measured at 225\,GHz}}
\end{table}


\begin{figure}[htbp] 
   \begin{center}
   \includegraphics[width=6in]{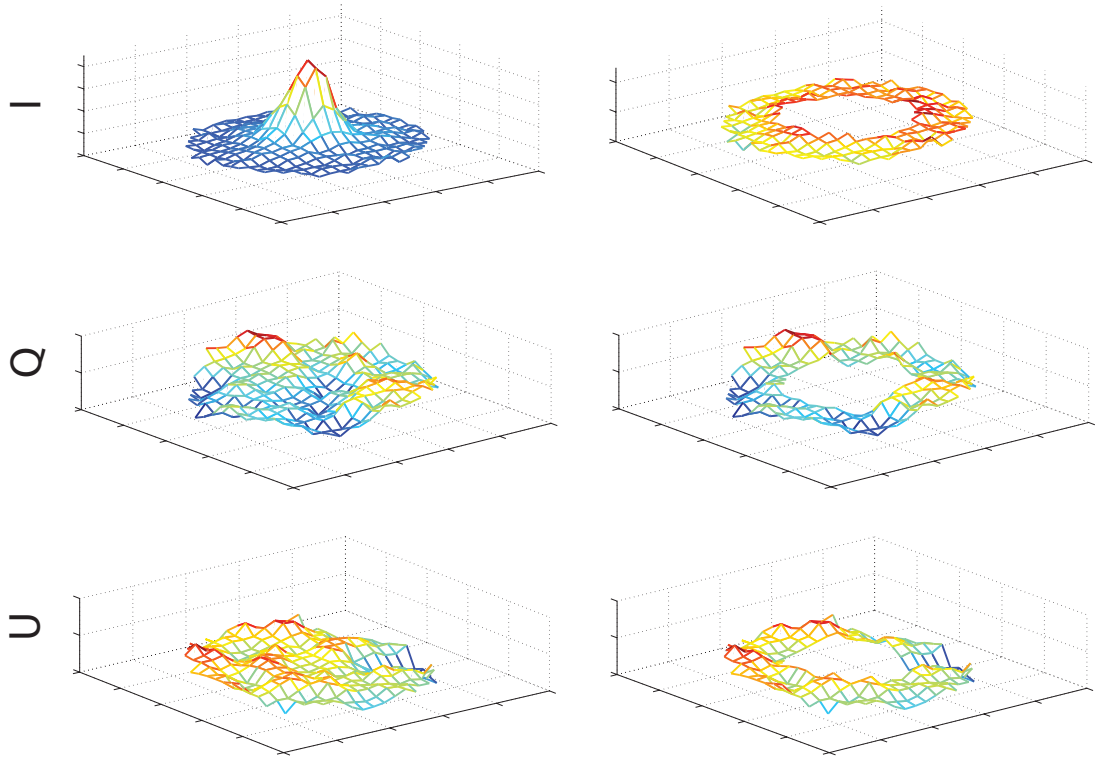} 
   \caption{\texttt{SHARPCOMBINE} output maps of $I$, $Q$, and $U$ for DG 
Tau.  Left panels show the full maps, with only the outermost high-noise 
regions removed, while the right panels show only the portion of each 
map that is used for computing the DC offset, i.e., the annulus regions. 
  The range of values encompassed by the vertical scale, in arbitrary 
flux units, is approximately -2.0 to +7.0 for the full $I$ map, -2.0 to 
+2.0 for the annulus $I$ map, and -0.2 to +0.2 for all other maps.
}
   \end{center}
   \label{fig:s_maps}
\end{figure}

\begin{figure}[htbp] 
   \centering
   \includegraphics[width=6in]{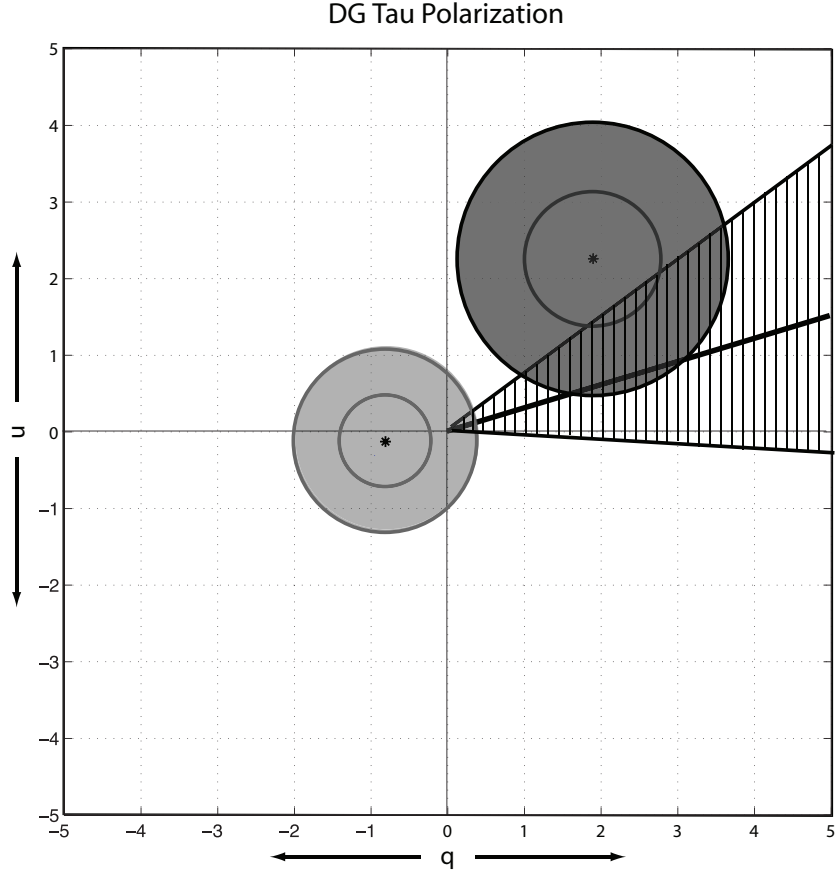} 
   \caption{Plot in Stokes space of \citet{tamura99} 850\,$\mu$m polarization measurement (dark grey circle, first quadrant) together with our 350\,$\mu$m measurement (light grey circle, third  quadrant).  Stokes $q$ are $u$ are plotted in percent.  Circles signify 1\,$\sigma$ and 2\,$\sigma$ errorbars.  The two measurements do not agree within 2\,$\sigma$, indicating structure in the polarization spectrum.  The vertically hatched region denotes the locus of points that are consistent with polarization orientated orthogonally to the plane of the disk.  (The disk is at position angle $99 \pm 10^{\circ}$.)  The 850\,$\mu$m point is consistent with polarization perpendicular to the disk plane; the 350\,$\mu$m measurement is consistent with zero.}
   \label{fig:dgtaupol}
\end{figure}

\bibliography{ms}

\begin{thebibliography}{28}
\expandafter\ifx\csname natexlab\endcsname\relax\def\natexlab#1{#1}\fi

\bibitem[{{Adams} {et~al.}(1990){Adams}, {Emerson}, \& {Fuller}}]{adams90}
{Adams}, F.~C., {Emerson}, J.~P., \& {Fuller}, G.~A. 1990, \apj, 357, 606

\bibitem[{{Beckwith} \& {Sargent}(1991)}]{beckwith91}
{Beckwith}, S.~V.~W., \& {Sargent}, A.~I. 1991, \apj, 381, 250

\bibitem[{{Beckwith} {et~al.}(1990){Beckwith}, {Sargent}, {Chini}, \&
  {Guesten}}]{beckwith90}
{Beckwith}, S.~V.~W., {Sargent}, A.~I., {Chini}, R.~S., \& {Guesten}, R. 1990,
  \aj, 99, 924

\bibitem[{{Cho} \& {Lazarian}(2007)}]{cho07}
{Cho}, J., \& {Lazarian}, A. 2007, \apj, 669, 1085

\bibitem[{{Cohen} \& {Kuhi}(1979)}]{cohen79}
{Cohen}, M., \& {Kuhi}, L.~V. 1979, \apjs, 41, 743

\bibitem[{{D'Alessio} {et~al.}(2001){D'Alessio}, {Calvet}, \&
  {Hartmann}}]{dalessio01}
{D'Alessio}, P., {Calvet}, N., \& {Hartmann}, L. 2001, \apj, 553, 321

\bibitem[{{Dowell}(1997)}]{dowell97}
{Dowell}, C.~D. 1997, \apj, 487, 237

\bibitem[{{Draine}(2006)}]{draine06}
{Draine}, B.~T. 2006, \apj, 636, 1114

\bibitem[{{Hall}(1949)}]{hall49}
{Hall}, J.~S. 1949, Science, 109, 166

\bibitem[{{Hildebrand}(1988)}]{hildebrand88}
{Hildebrand}, R.~H. 1988, \qjras, 29, 327

\bibitem[{{Hildebrand} {et~al.}(2000){Hildebrand}, {Davidson}, {Dotson},
  {Dowell}, {Novak}, \& {Vaillancourt}}]{hildebrand00}
{Hildebrand}, R.~H., {Davidson}, J.~A., {Dotson}, J.~L., {Dowell}, C.~D.,
  {Novak}, G., \& {Vaillancourt}, J.~E. 2000, \pasp, 112, 1215

\bibitem[{{Hiltner}(1949)}]{hiltner49}
{Hiltner}, W.~A. 1949, Science, 109, 165

\bibitem[{{Houde} \& {Vaillancourt}(2007)}]{houde07}
{Houde}, M., \& {Vaillancourt}, J.~E. 2007, \pasp, 119, 871

\bibitem[{{Howell}(2000)}]{howell00}
{Howell}, S.~B. 2000, {Handbook of CCD Astronomy} (Cambridge observing
  handbooks for research astronomers, New York : Cambridge University Press)

\bibitem[{{Jones} {et~al.}(1992){Jones}, {Klebe}, \& {Dickey}}]{jones92}
{Jones}, T.~J., {Klebe}, D., \& {Dickey}, J.~M. 1992, \apj, 389, 602

\bibitem[{{Kirby} {et~al.}(2005){Kirby}, {Davidson}, {Dotson}, {Dowell}, \&
  {Hildebrand}}]{kirby05}
{Kirby}, L., {Davidson}, J.~A., {Dotson}, J.~L., {Dowell}, C.~D., \&
  {Hildebrand}, R.~H. 2005, \pasp, 117, 991

\bibitem[{{Kitamura} {et~al.}(1996){Kitamura}, {Kawabe}, \&
  {Saito}}]{kitamura96}
{Kitamura}, Y., {Kawabe}, R., \& {Saito}, M. 1996, \apjl, 465, L137

\bibitem[{{Krejny}(2008)}]{krejny08}
{Krejny}, M. 2008, PhD thesis, Northwestern University

\bibitem[{{Lazarian}(2007)}]{lazarian07}
{Lazarian}, A. 2007, \jqsrt, 106, 225

\bibitem[{{Li} {et~al.}(2008){Li}, {Dowell}, {Kirby}, {Novak}, \&
  {Vaillancourt}}]{li08}
{Li}, H., {Dowell}, C.~D., {Kirby}, L., {Novak}, G., \& {Vaillancourt}, J.~E.
  2008, \ao, 47, 422

\bibitem[{{Mannings} \& {Emerson}(1994)}]{mannings94}
{Mannings}, V., \& {Emerson}, J.~P. 1994, \mnras, 267, 361

\bibitem[{{Mundt} \& {Fried}(1983)}]{mundt83}
{Mundt}, R., \& {Fried}, J.~W. 1983, \apjl, 274, L83

\bibitem[{{Novak} {et~al.}(1989){Novak}, {Gonatas}, {Hildebrand}, {Platt}, \&
  {Dragovan}}]{novak89}
{Novak}, G., {Gonatas}, D.~P., {Hildebrand}, R.~H., {Platt}, S.~R., \&
  {Dragovan}, M. 1989, \apj, 345, 802

\bibitem[{{Rodmann} {et~al.}(2006){Rodmann}, {Henning}, {Chandler}, {Mundy}, \&
  {Wilner}}]{rodmann06}
{Rodmann}, J., {Henning}, T., {Chandler}, C.~J., {Mundy}, L.~G., \& {Wilner},
  D.~J. 2006, \aap, 446, 211

\bibitem[{{Sargent} \& {Beckwith}(1989)}]{sargent89}
{Sargent}, A.~I., \& {Beckwith}, S.~V.~W. 1989, in Lecture Notes in Physics,
  Berlin Springer Verlag, Vol. 350, IAU Colloq. 120: Structure and Dynamics of
  the Interstellar Medium, ed. G.~{Tenorio-Tagle}, M.~{Moles}, \& J.~{Melnick},
  215

\bibitem[{{Stahler} \& {Palla}(2004)}]{stahler04}
{Stahler}, S.~W., \& {Palla}, F. 2004, {The Formation of Stars} (Weinheim:
  Wiley-VCH)

\bibitem[{{Tamura} {et~al.}(1999){Tamura}, {Hough}, {Greaves}, {Morino},
  {Chrysostomou}, {Holland}, \& {Momose}}]{tamura99}
{Tamura}, M., {Hough}, J.~H., {Greaves}, J.~S., {Morino}, J.-I.,
  {Chrysostomou}, A., {Holland}, W.~S., \& {Momose}, M. 1999, \apj, 525, 832

\bibitem[{{Werner} {et~al.}(1983){Werner}, {Capps}, \& {Dinerstein}}]{werner83}
{Werner}, M.~W., {Capps}, R.~W., \& {Dinerstein}, H.~L. 1983, \apjl, 265, L13

\end{thebibliography}
\bibliographystyle{apj}

\end{document}